%% file: transverseflow.tex
\documentclass[12pt,a4paper,twocolumn]{narms}

\usepackage{subfigure} 
\usepackage{psfig}
\usepackage{timesmt}   
\usepackage{chicaco}   

\usepackage{graphicx}

\makeatletter
\renewcommand{\section}{\@startsection%
{section}%
{1}%
{0mm}%
{- \baselineskip}%
{0.15\baselineskip}%
{\normalfont\normalsize}}%

\renewcommand{\subsection}{\@startsection
{subsection}%
{2}%
{0mm}%
{-\baselineskip}%
{0.15\baselineskip}%
{\normalfont\normalsize}}%
\makeatother

\def\anglefitA{0.208}
\def\anglefitB{5.85}

\def\posfitchi2{1.5}

\def\parfitchi2{4.3}

\def\anglefitposchi2{0.36}

\begin{document}

\title{Flow separation in the lee of transverse dunes}

\author{\large {Volker Schatz and Hans J. Herrmann}\\
{\em Institute for Computational Physics, Stuttgart University, Germany}}

\abstract{We investigate flow separation in the air flow over transverse sand
  dunes.  CFD simulations of the air flow over differently shaped dunes are
  performed.  The length of the recirculation region after the brink of the
  dune is found to depend strongly on the shape of the dune.  We find that the
  nondimensionalised separation length depends linearly on the slope of the
  dune at the brink within a large interval.  A phenomenological expression for
  the separation length is given.}

\maketitle

\section{INTRODUCTION}

Dunes are naturally occurring, beautifully shaped sand deposits.  Since the
middle of the previous century, they have attracted the attention of scientists
who have been seeking to model them and understand the processes leading to
their formation.  From the point of view of the physicist, sand dunes
constitute a variable boundary problem: The air flow is determined by the shape
of the dune and in turn influences the dune shape by transporting sand grains.
Therefore the air flow over dunes is of great importance for understanding dune
formation and evolution.

Since the start of scientific interest in dunes, there has been some work on
this topic, both theoretical~\cite{NelsonSmith89,Parsons04} and
experimental~\cite{Engel81,SweetKocurek90}.  However, due to the difficult
nature of the problem, these works have only tackled part of the problem.  In
several publications, transverse dunes have been modeled as triangular
structures~\cite{Engel81,Parsons04,Parsons04a}.  Field measurements of air flow
over dunes tend to lack measurements of the dune
profile~\cite{SweetKocurek90,FrankKocurek96}.

A recent field measurement~\cite{Parteli04} suggests that the shape of
transverse dunes has significant influence on the length of the recirculation
region.  Since the sand transport in the recirculation region in the lee of a
dune is negligible, the foot of the following dune shape is located at or
downwind of the flow reattachment point.  Therefore the distance of closely
spaced dunes is a measure of the length of the recirculation region.
In~\cite{Parteli04} the separation length after different dunes was determined
in this way.

In this work we will present results for widely spaced or isolated transverse
dunes.  This is to some extent an idealisation.  However, we think this is
valid and useful: We want to concentrate on the effect of the dune shape, other
things being equal, and the presence and shape of neighbouring dunes would
constitute additional parameters.


\section{METHOD}

Our simulations were performed with the computational fluid dynamics software
FLUENT.  The wind flow over dunes is fully turbulent.  We simulated the
Reynolds-averaged Navier-Stokes equations using the $k$-$\epsilon$ model for
closing the equations.

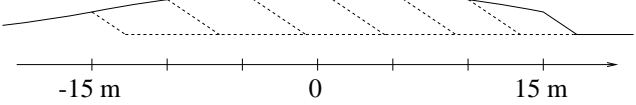
\begin{figure}
\input{shape.pstex_t}
\caption{The seven different dune shapes investigated.  The scale displays the 
  brink position.  The crest height of the dunes with negative brink position
  equals the brink height, the height of those with positive brink position is
  3 metres.}
\label{fig:shape}
\end{figure}

The simulations were two-dimensional, implying a wind direction perpendicular
to the dunes.  The cross sections of the dune shapes were constructed from two
circle segments, a concave one modeling the foot of the dune and a convex one
for the crest.  We verified that this is realistic enough to describe the
transverse dune shapes measured in \cite{Parteli04} well.  To obtain different
shapes, the position of the slip face was varied from the start to the end of
the convex part, see Figure~\ref{fig:shape}.  Note that this has the
consequence that not all the dunes have the same height.  The respective
heights and other geometrical data are given in Table~\ref{tab:results}.

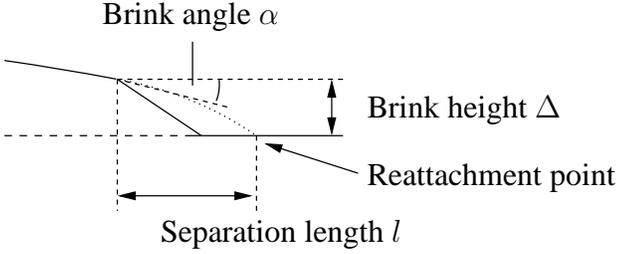
\begin{figure}
\input{geometry.pstex_t}
\caption{The geometric variables characterising the dune shapes.  The brink angle is positive for dunes with a sharp brink and negative for round dunes as the one shown in the figure.}
\label{fig:geometry}
\end{figure}

\begin{table*}
{\offinterlineskip
\def\tfil{\hskip 11pt plus 1fil \relax}
\halign{\vrule height2.5ex depth1ex width 0.7pt \tfil#\qquad&\vrule\quad \tfil#&#\tfil 
&\vrule\quad \tfil#&#\tfil &\vrule\quad \tfil#&#\tfil &\vrule\quad \tfil#&#\tfil 
&\vrule\  \tfil#&#\tfil &\vrule\tfil#&#\tfil \vrule width 0.7pt\cr
\noalign{\hrule height 0.7pt}
\omit\vrule height2.5ex depth 0.65ex width 0.7pt\tfil Brink pos. \tfil\ 
&\omit\span\omit\vrule\tfil Height  \tfil 
&\omit\span\omit\vrule\tfil Brink  \tfil 
&\omit\span\omit\vrule\tfil Angle at  \tfil
&\omit\span\omit\vrule\tfil Separation  \tfil 
&\omit\span\omit\vrule\tfil Error  \tfil 
&\omit\span\omit\vrule\tfil $l/\Delta$ \tfil \vrule width 0.7pt \cr
\omit\vrule height2.15ex depth 1ex width 0.7pt\tfil $d$ [m] \tfil\ 
&\omit\span\omit\vrule\tfil $H$ [m]  \tfil 
&\omit\span\omit\vrule\tfil height $\Delta$ [m]  \tfil 
&\omit\span\omit\vrule\tfil brink $\alpha$ [$^\circ$]  \tfil
&\omit\span\omit\vrule\tfil length $l$ [m]  \tfil 
&\omit\span\omit\vrule\tfil estimate [m]  \tfil 
&\omit\span\omit\vrule\tfil  \vrule width 0.7pt \cr
\noalign{\hrule height 0.7pt}
-15     & 1&.5   & 1&.5   &  11&.4  & 11&.74  & $\pm$0&.026$\;\pm 0.5$ & 7&.83 \cr \noalign{\hrule}
-10     & 2&.337 & 2&.337 &   7&.6  & 17&.33  & $\pm$0&.048$\;\pm 0.5$ & 7&.41 \cr \noalign{\hrule}
-5      & 2&.835 & 2&.835 &   3&.78 & 18&.88  & $\pm$0&.042$\;\pm 0.5$ & 6&.66 \cr \noalign{\hrule}
0       & 3&     & 3&     &   0&    & 17&.30  & $\pm$0&.065$\;\pm 0.5$ & 5&.77 \cr \noalign{\hrule}
5       & 3&     & 2&.835 &  -3&.78 & 14&.44  & $\pm$0&.050$\;\pm 0.5$ & 5&.09 \cr \noalign{\hrule}
10      & 3&     & 2&.337 &  -7&.6  &  9&.97  & $\pm$0&.041$\;\pm 0.5$ & 4&.27 \cr \noalign{\hrule}
15      & 3&     & 1&.5   & -11&.4  &  4&.96  & $\pm$0&.021$\;\pm 0.5$ & 3&.31 \cr \noalign{\hrule height 0.7pt}
}}
\vskip 3pt
\caption{Geometry of the simulated dunes and results for length of flow separation.  See Figure~\protect{\ref{fig:geometry}} for a definition of the geometric variables.  The brink angle is defined to be positive if the upwind slope is positive at the brink.  The first error in the separation length is the statistical error in the determination of the length from the simulation data, the second error is the systematic error of the simulation (see text).}
\label{tab:results}
\end{table*}

The velocity profile at the influx boundary of the simulation region was set to
the logarithmic profile which forms in flow over a plane in neutral atmospheric
conditions.  The shear velocity was chosen to be 0.4 m/s, the roughness length
$25 \mu$m.  The size of the roughness elements on the ground, the grain size,
was set to $250 \mu$m.  The roughness length is generally considered to be of
the order of the grain size or up to a factor of 30 smaller \cite{Wright97}.

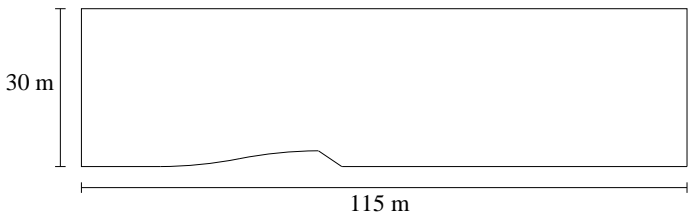
\begin{figure}[h]
\input{simregion.pstex_t}
\caption{The simulated region around the dune.}
\label{fig:simregion}
\end{figure}

The region around the dune in which the flow was simulated was chosen large
enough so that the boundaries did not influence the results.  This was verified
by choosing larger simulations areas for some dune shapes and comparing the
results.  From the brink position 0, the simulation region extends 45 m to the
left and 70 m to the right (see Figure~\ref{fig:simregion}).  The height of the
simulated region was chosen to be 30 m for all dunes except the one with brink
position -15 m, where 20 m was found to be sufficient.

The length of flow separation, our quantity of interest, was found to depend
slightly on the spacing of the simulation grid.  Therefore we performed the
simulation of the flow over each dune with three different grids and
interpolated the separation lengths to the continuum.  The average grid
spacings were 10, 7 and 5~cm, respectively.

\section{RESULTS}

Table~\ref{tab:results} shows the results for all dune shapes.  The length of
flow separation, our quantity of interest, was measured from the slip face
brink, where the flow separates, to the flow reattachment point (see
Figure~\ref{fig:geometry}).  The errors were estimated to be one grid spacing
for the determination of the flow reattachment point.  The separation lengths
determined for the different grids and their errors were interpolated to the
continuum with the standard linear regression formulas.  To the statistical
error we added quadratically a systematic error of 0.5 metres.  The systematic
error accounts for biases which may be inherent in the turbulence model and
parameter settings used.  We expect it to be strongly correlated.

\begin{figure}
\input{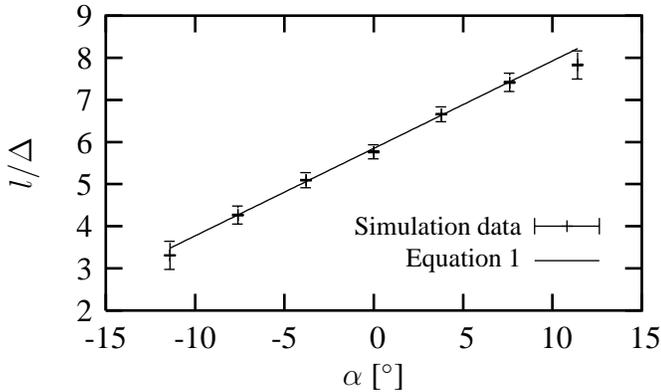}
\caption{Dependence of the nondimensionalised separation length on the 
  angle~$\alpha$.  The relationship is remarkably linear.  The rightmost value
  of $\alpha$ belongs to the dune with the sharpest brink.}
\label{fig:brinkangle}
\end{figure}

To nondimensionalise the separation length $l$, it was divided by the height of
the slip face.  This dimensionless quantity is universal, that is it does not
depend on the absolute height of the dune.  That is because the flow is already
fully turbulent for small dunes.  We find that $l/\Delta$ is larger for dunes
with a sharp brink than for rounded dunes.  It depends linearly on the brink
position~$d$ respectively the angle of the dune shape at the brink,~$\alpha$.
As can be seen in Figure~\ref{fig:brinkangle}, the linear relation extends up
to an absolute angle of~$7.5^\circ$.  This amounts to a variation of the dune
length between 6.7 and 13.3 times the height.  Fitting the relation
\begin{equation}
l(\alpha)/\Delta(\alpha)= A\cdot \alpha + B\,,
\label{eq:brinkangle}
\end{equation}
we obtain $A=\anglefitA/^\circ$ and $B=\anglefitB$.  The points with brink position
$\pm 15$ were ignored for this fit since they deviate from the linear law.

\begin{figure}
\input{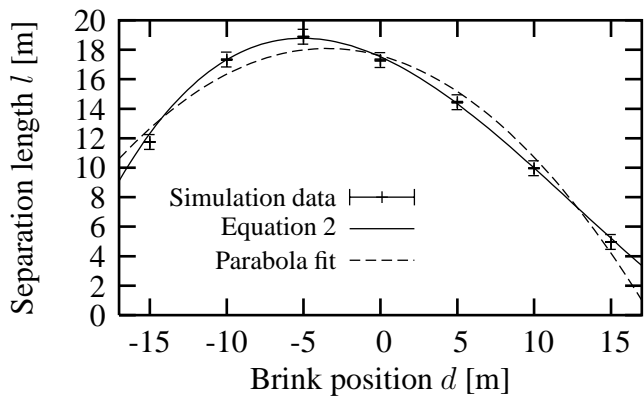}
\caption{Dependence of the flow separation length on the brink position.
        The expression derived from the linear angle dependence displayed in
        Figure~\ref{fig:brinkangle} provides a much better fit than a
        parabola.}
\label{fig:brinkpos}
\end{figure}

Since the brink angle $\alpha$ and the brink position $d$ are related by the
geometry of the dune, the fit (\ref{eq:brinkangle}) can be reformulated to give
the separation length in terms of the brink position:
\begin{eqnarray}
\label{eq:anglefitpos}
l(\alpha(d))&=& (A\cdot\alpha(d) + B)\;\Delta(\alpha(d))
\nonumber\\
&=& \left(-A \,\arcsin \frac dR + B\right)\cdot{} \\ 
&&{}\cdot\left(H_{\hbox{\scriptsize max}} 
- d\,\tan \left( \frac12 \,\arcsin\frac dR \right) \right)
\nonumber
\end{eqnarray}
This equation contains the height of the round dunes, $H_{\hbox{\scriptsize
    max}}=3\;$m, and the radius of the circle segments used to model the shape,
$R=75.75\;$m.  Remarkably, this expression fits significantly better than a
parabola (see Figure~\ref{fig:brinkpos}) even though it has a smaller number of
parameters.  It has a $\chi^2=\anglefitposchi2$ compared to
$\chi^2=\parfitchi2$ for the parabola fit.

Equation~\ref{eq:anglefitpos} can be approximated by using $\arcsin x\approx x$
and $\tan x \approx x$.  In this way one obtains a polynomial expression for
the dependence of the separation length on the brink position:
\begin{eqnarray}
l(d)&=& \left( -A\frac dR +B \right)
        \left(H_{\hbox{\scriptsize max}}-\frac{d^2}{2\,R}\right)
\nonumber\\
&=& \frac{A}{2\,R^2}\;d^3 - \frac B{2\,R}\;d^2 
    - \frac{A\,H_{\hbox{\scriptsize max}}}R\;d + B\,H_{\hbox{\scriptsize max}}
\nonumber\\
&=& C\;d^3 + D\;d^2 + E\;d + F
\label{eq:polyfit}
\end{eqnarray}
The coefficients are: $C=0.00103\;$m$^{-2}$, $D=0.0386\;$m$^{-1}$, $E=0.472$,
and $F=17.55\;$m.  The values of this expression are indistinguishable from
(\ref{eq:anglefitpos}) in the range of $d$ which we investigated.  Therefore we
do not plot it separately in Figure~\ref{fig:brinkpos}.

\section{DISCUSSION}

Comparison of our results to other work is hampered by the fact that the
dependence of flow separation on the dune shape has not been investigated
before.  Therefore, the following comparisons are to be understood as
consistency checks.

A recent review of air flow over transverse dunes \cite{WalkerNickling02} cites
values of 4--10 for $l/\Delta$.  Our results also lie within that range (see
Figure~\ref{fig:brinkangle}).  Engel \cite{Engel81} finds values for the
nondimensionalised separation length between 4 and a little over 6, depending
on the roughness and the aspect ratio of triangular dunes.  In \cite{Parsons04}
a wide range of between 3 and 15 is given for the same quantity.  The values
for an aspect ratio of 0.1, which applies to our dune with $\alpha=0$, are 5.67
and 8.13, depending on the height.  This compares well with our value of 5.76.
The discrepancy can be explained by the different shape, in particular the fact
that our dune shape for $\alpha=0$ has a horizontal tangent at the brink,
whereas the dunes in \cite{Parsons04} are triangular.

Unlike our simulations, which treated an isolated dune shape, the field
measurements \cite{Parteli04} were performed in a closely spaced dune field.
The authors find that the distance between the brink of each dune and the foot
of the following one is typically four times the height or below.  Under the
assumption that the dune field is stationary, this distance is a measure of the
separation length.  This is at the lower end of the separation lengths we
obtain, for very rounded dunes.  The dunes with the smallest separation length
in \cite{Parteli04} are indeed very round.  The small values for other dunes
can be put down to the different situation of closely spaced transverse dunes.

\section{CONCLUSIONS}

We have determined the length of flow separation in the lee of isolated
transverse dunes of different shapes.  The separation length nondimensionalised
by division by the slip face height was found to be larger for dunes with a
sharper brink.  For a wide range of dune shapes, this growth is well described
by the linear relationship~(\ref{eq:brinkangle}).

The dependence of the absolute separation length on the position of the slip
face is described by the expression~(\ref{eq:anglefitpos}) more accurately than
by a parabola.  The maximal separation length does not occur for dune shapes
with a horizontal tangent at the brink, but for shapes with a somewhat sharper
brink.

These results were obtained for isolated dunes up to 3 metres in height.  It
would be interesting to see how the flow separation length is influenced by the
presence of other dunes nearby, as in a closely spaced dune field.  This topic
will be treated in a future publication.

\bibliography{dune}

\end{document}

%% file: shape.pstex_t
\begin{picture}(0,0)%
\includegraphics{shape.pstex}%
\end{picture}%
\setlength{\unitlength}{829sp}%
\begingroup\makeatletter\ifx\SetFigFont\undefined%
\gdef\SetFigFont#1#2#3#4#5{%
  \reset@font\fontsize{#1}{#2pt}%
  \fontfamily{#3}\fontseries{#4}\fontshape{#5}%
  \selectfont}%
\fi\endgroup%
\begin{picture}(18937,3172)(4036,-9961)
\put(13501,-9961){\makebox(0,0)[b]{\smash{\SetFigFont{10}{12.0}{\rmdefault}{\mddefault}{\updefault}\special{ps: gsave 0 0 0 setrgbcolor}0\special{ps: grestore}}}}
\put(20251,-9961){\makebox(0,0)[b]{\smash{\SetFigFont{10}{12.0}{\rmdefault}{\mddefault}{\updefault}\special{ps: gsave 0 0 0 setrgbcolor}15 m\special{ps: grestore}}}}
\put(6751,-9961){\makebox(0,0)[b]{\smash{\SetFigFont{10}{12.0}{\rmdefault}{\mddefault}{\updefault}\special{ps: gsave 0 0 0 setrgbcolor}-15 m\special{ps: grestore}}}}
\end{picture}

%% file: geometry.pstex_t
\begin{picture}(0,0)%
\includegraphics{geometry.pstex}%
\end{picture}%
\setlength{\unitlength}{2072sp}%
\begingroup\makeatletter\ifx\SetFigFont\undefined%
\gdef\SetFigFont#1#2#3#4#5{%
  \reset@font\fontsize{#1}{#2pt}%
  \fontfamily{#3}\fontseries{#4}\fontshape{#5}%
  \selectfont}%
\fi\endgroup%
\begin{picture}(4387,3022)(18879,-9533)
\put(23266,-7936){\makebox(0,0)[lb]{\smash{\SetFigFont{12}{14.4}{\rmdefault}{\mddefault}{\updefault}\special{ps: gsave 0 0 0 setrgbcolor}Brink height $\Delta$\special{ps: grestore}}}}
\put(23266,-8746){\makebox(0,0)[lb]{\smash{\SetFigFont{12}{14.4}{\rmdefault}{\mddefault}{\updefault}\special{ps: gsave 0 0 0 setrgbcolor}Reattachment point\special{ps: grestore}}}}
\put(20791,-9421){\makebox(0,0)[lb]{\smash{\SetFigFont{12}{14.4}{\rmdefault}{\mddefault}{\updefault}\special{ps: gsave 0 0 0 setrgbcolor}Separation length $l$\special{ps: grestore}}}}
\put(21151,-6811){\makebox(0,0)[b]{\smash{\SetFigFont{12}{14.4}{\rmdefault}{\mddefault}{\updefault}\special{ps: gsave 0 0 0 setrgbcolor}Brink angle $\alpha$\special{ps: grestore}}}}
\end{picture}

%% file: simregion.pstex_t
\begin{picture}(0,0)%
\includegraphics{simregion.pstex}%
\end{picture}%
\setlength{\unitlength}{580sp}%
\begingroup\makeatletter\ifx\SetFigFont\undefined%
\gdef\SetFigFont#1#2#3#4#5{%
  \reset@font\fontsize{#1}{#2pt}%
  \fontfamily{#3}\fontseries{#4}\fontshape{#5}%
  \selectfont}%
\fi\endgroup%
\begin{picture}(29058,8583)(4726,-9511)
\put(20701,-9511){\makebox(0,0)[b]{\smash{\SetFigFont{9}{10.8}{\rmdefault}{\mddefault}{\updefault}\special{ps: gsave 0 0 0 setrgbcolor}115 m\special{ps: grestore}}}}
\put(4726,-4336){\makebox(0,0)[lb]{\smash{\SetFigFont{9}{10.8}{\rmdefault}{\mddefault}{\updefault}\special{ps: gsave 0 0 0 setrgbcolor}30 m\special{ps: grestore}}}}
\end{picture}